\documentclass{PoS}
\usepackage{epsfig}

\PoS{PoS(LAT2005)294}

\title{Studying glueball masses in non-Abelian LGT with the LW algorithm }

\ShortTitle{Studying glueball masses in non-Abelian LGT with the LW algorithm }

\author{\speaker{Marco Panero}
\\

        Dublin Institute for Advanced Studies\\

        E-mail: \email{panero@stp.dias.ie}}

\abstract{We address a study of glueball masses in the confining regime of $SU(2)$ in $D=3$ using an algorithm inspired by the multi-level scheme. Our method, which exploits the locality of the action to achieve high precision results, is based on a technique already used for compact QED, and generalises it to the non-Abelian case. We discuss the main features of this method, in comparison with other algorithms that have been used in similar studies.}

\FullConference{XXIIIrd International Symposium on Lattice Field Theory\\

                 25-30 July 2005\\

                 Trinity College, Dublin, Ireland}

\usepackage{amssymb}
\usepackage{latexsym,cite}
\newcommand{\eq}{\begin{equation}}
\newcommand{\en}{\end{equation}}
\newcommand{\qe}{\end{equation}}
\newcommand{\ear}{\begin{eqnarray}}
\newcommand{\eqa}{\begin{eqnarray}}
\newcommand{\rae}{\end{eqnarray}}
\newcommand{\ena}{\end{eqnarray}}
\newcommand{\beq}{\begin{equation}}
\newcommand{\eeq}{\end{equation}}
\newcommand{\bea}{\begin{eqnarray}}
\newcommand{\eea}{\end{eqnarray}}

\begin{document}

\section{Introduction}

The basic idea underlying lattice calculations of the glueball spectrum in pure gauge theory is very simple: given a zero-momentum lattice operator $\phi$ with quantum numbers $J^{PC}$, the connected correlator at large time separations $\Delta t$ is expected to behave like:
\eq
\label{naivelarget}
\langle \phi^\dagger(t_0+\Delta t) \phi(t_0) \rangle_{\mbox{\tiny{conn}}} \simeq |\langle \mbox{vac}|\phi|0\rangle|^2 \; e^{ - E_0 \Delta t } 
\;\;\; ,
\en
where $E_0$ is the energy of the lightest state $| 0 \rangle$ that couples to $\phi$. The operator must have a sufficiently large overlap with the wave-functional associated to the required channel, and must be sufficiently smooth over the physical length scales characterising the wave-functional itself; this can be achieved by means of standard variational methods \cite{Berg:1982kp,Luscher:1990ck} and through ``smearing'' \cite{Albanese:1987ds} and ``blocking'' \cite{Teper:1987wt} techniques, which allow to obtain precise results for $SU(N)$ gauge theories \cite{Teper:1998te,Lucini:2001ej,Lucini:2002wg,Lucini:2004my}. However, these calculations are technically non-trivial, due to the fact that the connected correlators are typically very small quantities, affected by large relative errors at the time-like separations where one expects to extract a signal for the mass; a solution was proposed in \cite{Morningstar:1997ff,Morningstar:1999rf}, using anisotropic lattices with a fine spacing in the time-like direction, in combination with an improved action.

In this work we discuss another method which is expected to improve the result precision in measurements of glueball correlation functions: following the idea proposed in \cite{Luscher:2001up}, we use a multi-level algorithm to calculate the observable expectation values from products of sublattice averages; in particular, for the scalar channel, we consider a method which involves (discretised) time-like derivatives of the disconnected correlator. Since this method has been successfully used for compact QED \cite{Majumdar:2003xm}, we are encouraged to test it in a non-Abelian model, focusing on $SU(2)$ lattice gauge theory in $2+1$ dimensions. Another study of the Yang-Mills spectrum using the multi-level algorithm can be found in \cite{Meyer:2002cd,Meyer:2003hy}.

\section{The LW multi-level algorithm \dots}
 
L\"uscher and Weisz \cite{Luscher:2001up} proposed a multi-level algorithm which achieves exponential error reduction in measurements of Polyakov loop expectation values in quenched LGT; given the two-point Polyakov loop correlation function:
\eq
G(r) \equiv \langle P^*(r) P(0) \rangle \;\;\; , \;\;\; \mbox{with: }P(x)=\mbox{tr}\left( \prod_{t} U_0(t,\vec{x}) \right)
\;\;\; ,
\en
$P^*(r) P(0)$ can be factorized in a product of two-link operators $\mathbb{T}(t)$:
\eq
P^*(r) P(0) = \left\{ \prod_{t} \mathbb{T} (t) \right\}_{\alpha\alpha\gamma\gamma} \;\;\; , \;\;\; \mbox{with: } \mathbb{T}(t)_{\alpha\beta\gamma\delta} = U_0(t,\vec{r})^*_{\alpha\beta} U_0(t,\vec{0})_{\gamma\delta}
\;\;\; .
\en
whose expectation values can be estimated to a high degree of precision from sublattice averages --- conventionally denoted with the $[\dots]$ notation --- in time-like slabs of given thickness: the LW algorithm thus amounts to a variant of the multi-hit method \cite{Parisi:1983hm}, applied to pair of links; the scheme can be generalised to several nested sublevels. Due to locality of the gauge action, the different sublattices are independent of each other, provided the space-like link variables at their boundaries are kept fixed; the thickness of the sublattices must be such that they are still in the confined phase. The value of the $ P^*(r) P(0) $ product in the full lattice is then obtained multiplying the sublattice averages of the $\mathbb{T}$'s, and finally the two-point Polyakov loop correlator can be evaluated averaging over a sample of full-lattice configurations: in practice, after every (standard) update of the whole lattice, the algorithm performs a cycle of updates in the nested sublattices, keeping their boundaries fixed, to measure the $[ \mathbb{T}(t) \mathbb{T}(t+a) \dots ]$ sublattice averages.

The LW algorithm increases the result precision by damping the uncertainty due to short-distance fluctuations, through the sublevel averages: in a theory with non-vanishing mass gap, the algorithm leads to exponential error reduction with respect to standard simulations; if (at the lowest level) the whole lattice is subdivided in $n_s$ time-like slabs, then from $n$ measurements in each slab (which require an amount of computer-time comparable with $n$ full-lattice measurements in the standard approach) one effectively achieves a precision corresponding to $n^{n_s}$ measurements in the whole lattice. Clearly, the efficiency of the algorithm depends on some parameters which can be tuned by optimisation.

\section{\dots adapted to study the glueball spectrum}

To generalise the LW algorithm idea to the study of glueballs, the correlator in eq.~(\ref{naivelarget}) has to be expressed in a factorized form, in terms of quantities that can be measured to a high degree of precision from sublattice averages; the most obvious choice is to consider the zero-momentum component of $\phi$, evaluated in two time-slices belonging to different slabs. However, for the $0^{++}$ channel one has then to subtract the non-vanishing VEV of the operators, keeping into account only the measurements that have been used in the two-point function, as it was remarked in \cite{Meyer:2003hy}. Here, we follow a slightly different strategy, first used in the case of compact $U(1)$ LGT in four space-time dimensions \cite{Majumdar:2003xm}, considering time-derivatives of the full correlator:
\eq
\frac{\partial}{\partial t_1} \frac{\partial}{\partial t_0} \langle \phi^\dagger (t_1) \phi(t_0) \rangle \simeq  - \alpha m^2 \left( e^{-m(t_1-t_0)} + e^{-m(N_t-t_1+t_0)}  \right)
\;\;\; .
\en
On the lattice, the expression above can be approximated in terms of finite differences, to read:
\eq
\label{ddd}
F (t_1 -t_0) \equiv \nabla^f_{t_1} \nabla^f_{t_0}  \langle \phi^\dagger (t_1) \phi(t_0) \rangle \simeq 2 \alpha \left\{e^{-m(t_1-t_0)} + e^{-m(N_t-t_1+t_0)} \right\} \left( 1 - \cosh m \right)
\;\;\; .
\en
Then we can use the LW algorithm to measure:
\eq
\label{twofactors}
\nabla^f_{t_1} \nabla^f_{t_0}  \langle \phi^\dagger (t_1) \phi(t_0) \rangle = \langle \left[ \phi^\dagger(t_1+1) - \phi^\dagger(t_1)\right]  \left[ \phi(t_0 +1) - \phi(t_0)\right] \rangle
\;\;\; .
\en
Notice that the quantity in eq.~(\ref{twofactors}) is expressed as a product of only two factors, whereas in the original formulation of the LW algorithm \cite{Luscher:2001up} the Polyakov line correlator was completely factorized in a (generally) larger number of factors.

\section{Algorithm details}

We consider a cubic lattice with sizes $N_x=N_y=N_t=48$, and use the standard plaquette action:
\eq
S=\beta \sum_{\Box} \left( 1- \frac{1}{2} \Re \mbox{tr} U_\Box \right)\;\;\; , \;\;\; \mbox{with: } \beta=\frac{4}{ag^2}
\;\;\; ;
\en
the configurations are updated using a combination of the Kennedy-Pendleton heat-bath \cite{Kennedy:1985nu} followed by three overrelaxation \cite{Brown:1987rr} steps. 

The results discussed here have been obtained from simulations at $\beta=9.0$, which corresponds to a lattice spacing $a\simeq 0.074 $ fm \cite{Caselle:2004er}; in the measurements, we focused on the $J^{PC}=0^{++}$ channel, considering a family of operators obtained from rectangles with sizes ranging from 1 to 4, built from ``smeared'' variables \cite{Albanese:1987ds}.

In the practical implementation of the multi-level algorithm, one needs to tune some parameters according to optimisation criteria (see \cite{Luscher:2001up}), including the number of nested average sublevels, the thickness of the slabs, and the number of measurements in the sublattices. As it was pointed out in \cite{Luscher:2002qv}, it is quite remarkable that the LW algorithm does not necessarily require a large number of nested levels: for most applications, a two-level algorithm proves to be highly efficient (and our experience in the present study confirms this empirical observation); in any case, more complicated schemes, with many nested levels, can be implemented numerically in a simple, versatile and efficient way using recursive functions. An important parameter in the original LW algorithm is the thickness of the sublattices: for best efficiency, it has to be set to the minimum value compatible with the fact that the sublattice system is still in the confined phase, and a reasonable lower bound is half the inverse deconfinement temperature at the given value of $\beta$ \cite{Kratochvila:2003zj,Caselle:2004er}. Finally, a ``thumb-rule'' was suggested in \cite{Meyer:2002cd}, estimating the optimum number of sublattice measurements in a slab of width $L_{\mbox{\tiny{slab}}}$ to be of the order of $\exp \left(m_0 L_{\mbox{\tiny{slab}}} \right)$, where $m_0$ is the mass of the lightest state compatible with the quantum numbers of the operator which is measured.

In the preliminary results of the study that we present here, we kept the mentioned criteria in mind; a systematic study of the parameter choice is among the subjects of work in progress, and we refer the reader to a forthcoming, complete paper \cite{Panero_in_preparation} for a more exhaustive discussion.

\section{Preliminary results, observations, and comments}

\begin{figure}
\centerline{\includegraphics[height=100mm]{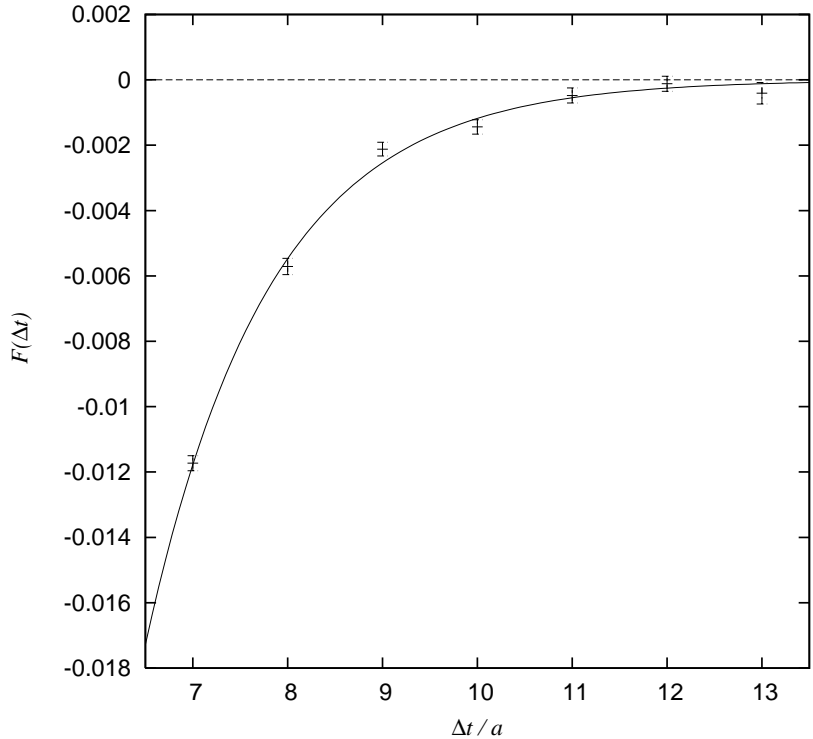}}
\vspace{1cm}
\caption{Results of the multi-level algorithm for the discretised double derivative of the correlator in the $0^{++}$ channel $F(\Delta t)$ \emph{versus} the time-slice separation in lattice units; the fitted mass reads: $ma=0.767(31)$.}
\label{plotfig}
\end{figure}

The information emerging from the preliminary results of this study shows that the multi-level algorithm improves the estimate of the lowest-lying glueball mass: we observed that the estimate for the lowest mass in the $0^{++}$ channel obtained from a fit to eq.~(\ref{ddd}) is affected by a smaller uncertainty, with respect to a standard algorithm requiring a similar computational effort (see table~\ref{comparisontab}); also, the multi-level algorithm allows to explore a larger range of time separations (see figure~\ref{plotfig}), where the signal obtained in standard simulations is already drowned in statistical noise.

\begin{table}
\centerline{\begin{tabular}{|c|c|c|}
\hline
Method & $am_{0^{++}}$ & $\chi^2_{\mbox{\tiny{red}}}$ \\ 
\hline
\hline
Standard algorithm & 0.802(58) & 1.29 \\
\hline
Multi-level algorithm & 0.767(31) & 0.57 \\
\hline
\end{tabular}}
\caption{Comparison of the estimates for the lowest-lying mass in the $0^{++}$ channel, obtained from runs requiring a similar amount of CPU time.}
\label{comparisontab}
\end{table}

We also observed that our choice of the parameters in the study of the $0^{++}$ channel has proved effective; in particular, the systematic error induced by the finite differences in eq.~(\ref{ddd}) did not bias the estimate of the lowest glueball mass with respect to the results available in literature \cite{Teper:1998te} --- at least within the (limited) precision of the preliminary study presented here.

Finally, we would like to point out that another advantage of using the multi-level algorithm in the study of the glueball spectrum (where, typically, one measures a large number of operators) is the reduced amount of data storage which is required, with respect to to standard simulations; this aspect, which was also remarked in \cite{Meyer:2003hy}, is simply due to the fact that the correlation matrices are built using the sublattice averages of the operators in a given channel, rather than the single measurements.

\section{Conclusions, perspectives and acknowledgements}

The analysis in progress is confirming that the LW multi-level algorithm can be successfully applied to numerical investigations of the glueball spectrum in non-Abelian lattice gauge theories: a conclusion in agreement with the results of similar studies.

In particular, we would like to highlight the following aspects:
\begin{itemize}
\item efficient error reduction with respect to standard simulations;
\item reduced data storage;
\item versatility: the algorithm can be used for a generic gauge theory, and it can be combined with other methods which improve the quality of the results.
\end{itemize}

A natural generalisation of our algorithm would be to use anisotropic lattices with a fine spacing along the time-like direction \cite{Morningstar:1997ff,Morningstar:1999rf}: this would enhance the data resolution at intermediate distances, with the further bonus of reducing the systematic error induced by the discretised time-derivatives used for the $0^{++}$ case.

Complete results of this work will be published in \cite{Panero_in_preparation}.

The author thanks Philippe de Forcrand, Ferdinando Gliozzi, Miho Koma, Yoshiaki Koma, Biagio Lucini and Antonio Rago for useful discussions, and acknowledges support received from Enterprise Ireland under the Basic Research Programme.

\end{document}